\documentclass[sigconf]{acmart}

\setcopyright{none}
\renewcommand\footnotetextcopyrightpermission[1]{}

\usepackage{fancyhdr}
\fancypagestyle{plain}{%
   \fancyhf{} %
   \fancyfoot[L]{ACM SIGCOMM Computer Communication Review}%
   \fancyfoot[R]{Volume Issue , 2022}%
}
\fancypagestyle{firstpagestyle}{%
   \fancyhf{} %
   \fancyfoot[L]{ACM SIGCOMM Computer Communication Review}%
   \fancyfoot[R]{Volume  Issue , 2022}%
}

\begin{teaserfigure}
	\parbox{\textwidth}{\centering\normalsize
		* Corresponding author: Aaron Ding (aaron.ding@tudelft.nl)
	}
	\vspace{10pt}
\end{teaserfigure}

\usepackage{balance}

\begin{document}

\title{Roadmap for Edge AI: A Dagstuhl Perspective}

\author{
\Large
Aaron Yi Ding\textsuperscript{1}*, Ella Peltonen\textsuperscript{2}, Tobias Meuser\textsuperscript{3}, Atakan Aral\textsuperscript{4}, Christian Becker\textsuperscript{5}, Schahram Dustdar\textsuperscript{6}, Thomas Hiessl\textsuperscript{6}, Dieter Kranzlm\"{u}ller\textsuperscript{7},
Madhusanka Liyanage\textsuperscript{8}, Setareh Magshudi\textsuperscript{9}, Nitinder Mohan\textsuperscript{10},\\J\"{o}rg Ott\textsuperscript{10}, Jan S. Rellermeyer\textsuperscript{11,}\textsuperscript{1}, Stefan Schulte\textsuperscript{12}, Henning Schulzrinne\textsuperscript{13}, G\"{u}rkan Solmaz\textsuperscript{14},\\ Sasu Tarkoma\textsuperscript{15}, Blesson Varghese\textsuperscript{16}, Lars Wolf\textsuperscript{17}\\
\vspace*{1mm}
\normalsize
\textsuperscript{1}TU Delft, \textsuperscript{2}University of Oulu, \textsuperscript{3}TU Darmstadt, \textsuperscript{4}University of Vienna,
\textsuperscript{5}University of Mannheim,
\textsuperscript{6}TU Wien, \textsuperscript{7}LMU Munich,
\textsuperscript{8}University College Dublin,
\textsuperscript{9}University of T\"{u}bingen,
\textsuperscript{10}TU Munich,
\textsuperscript{11}Leibniz University Hannover,
\textsuperscript{12}Hamburg University of Technology,
\textsuperscript{13}Columbia University,
\textsuperscript{14}NEC Lab,
\textsuperscript{15}University of Helsinki,
\textsuperscript{16}Queen's University Belfast,
\textsuperscript{17}TU Braunschweig
}

\begin{abstract}

Based on the collective input of Dagstuhl Seminar (21342), this paper presents a comprehensive discussion on AI methods and capabilities in the context of edge computing, referred as Edge AI. In a nutshell, we envision Edge AI to provide adaptation for data-driven applications, enhance network and radio access, and allow the creation, optimisation, and deployment of distributed AI/ML pipelines with given quality of experience, trust, security and privacy targets. The Edge AI community investigates novel ML methods for the edge computing environment, spanning multiple sub-fields of computer science, engineering and ICT. The goal is to share an envisioned roadmap that can bring together key actors and enablers to further advance the domain of Edge AI.

\end{abstract}


\maketitle

\section{Introduction}\label{sec:intro}

Edge computing promises to decentralise cloud applications while providing more bandwidth and reducing latency \cite{Varghese-IC2021}. These promises are delivered by moving application-specific computations between the cloud, the data producing devices, and the network infrastructure components at the edges of wireless and fixed networks~\cite{Mohan20:hotnets}. Meanwhile, the current Artificial Intelligence (AI) and Machine Learning (ML) methods assume computations are conducted in a powerful computational infrastructure \cite{EI-whitepaper2020}, such as datacenters with ample computing and data storage resources available. To shed light on the fast evolving domain that merges edge computing and AI/ML, referred as \textbf{Edge AI}, the recent Dagstuhl Seminar 21342\footnote{https://www.dagstuhl.de/en/program/calendar/semhp/?semnr=21342} has gathered community inputs from a diverse range of experts. The efforts result in this CCR paper that discusses both technical and societal demands for applying AI methods in the context of edge computing. 

As recent research and development go along, the 'Edge' itself remains a diffuse term. A commonly accepted definition of what the edge is, where it resides, and who provides it, is lacking across different communities and researchers\footnote{We deliberately renounce marketing-driven differentiation of edge vs. fog vs. mist computing in this work.}. Common understanding is shared about its properties: as compared to the cloud, its features are closeness (latency and topology), increased network capacity (effectively achievable data transmission rate), lower computational power, smaller scale, and higher heterogeneity of devices. Compared to the end devices (the final hop), it features increased computational and storage resources. It is an abstract entity to offload computation and storage without the detour to the cloud.

A raising area of tension arises from current AI and ML methods, which require powerful computational infrastructure \cite{EI-whitepaper2020} -- a demand that is better satisfied in a data center with ample available computing and data storage resources. However, sending the necessary raw data to the cloud puts pressure on the network w.r.t. bandwidth and throughput. Meanwhile, organisations are usually less keen on sharing (potentially restricted) data with commercial cloud providers. This tension is addressed by the fast evolving domain of Edge AI.

\begin{figure}
    \centering
    \includegraphics[width=\linewidth,trim=0 4.2cm 0 4.45cm, clip]{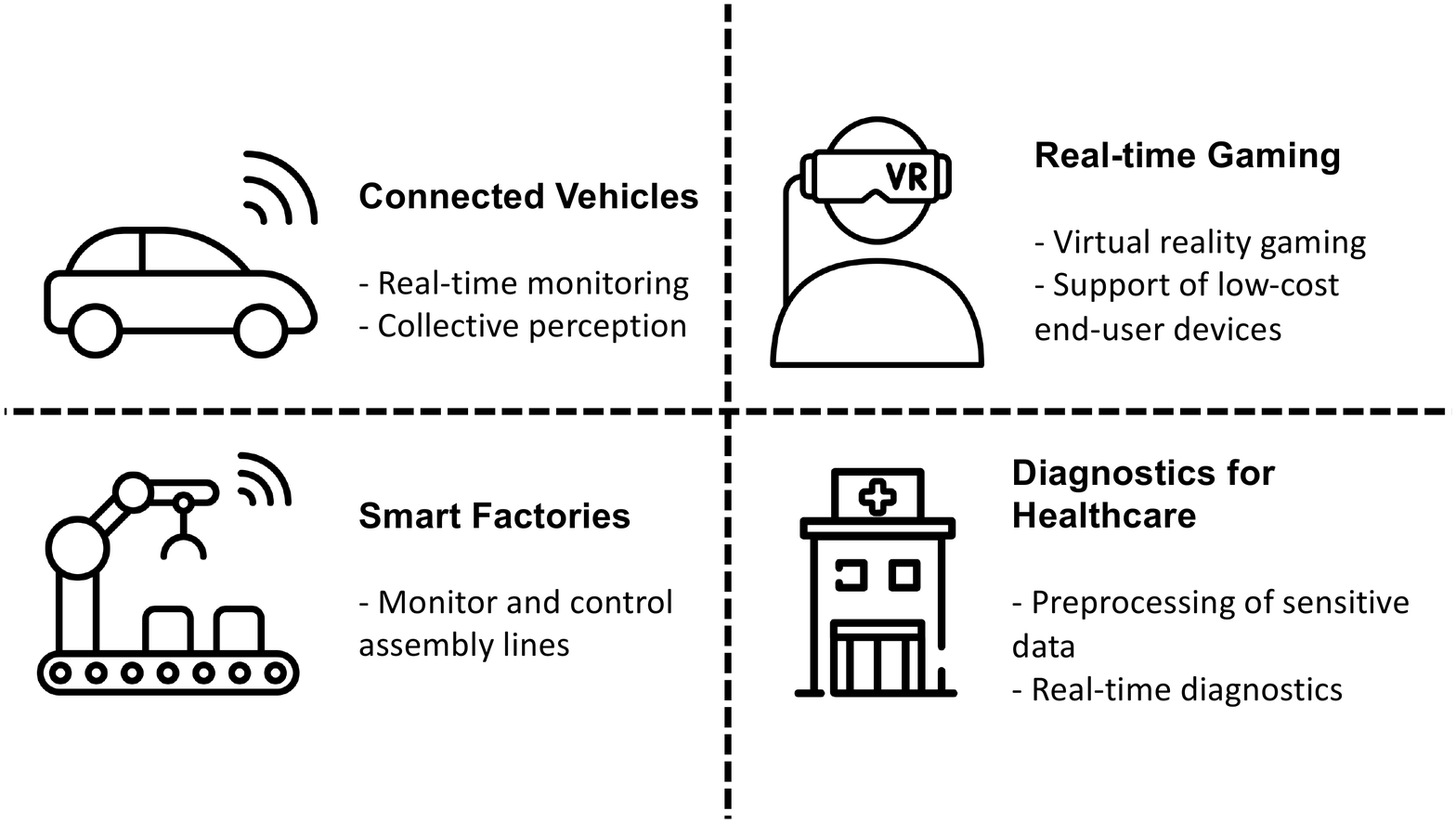}
    \caption{Usecases of Edge AI}
    \label{fig:usecases}
\end{figure}

As highlighted in Figure~\ref{fig:usecases}, Edge AI has gradually found its way to mainstream service domains such as connected vehicles, real-time gaming, smart factories, and healthcare. From infrastructure perspective, edge environments provide a unique layer for AI and also offer opportunities for existing technologies such as embedded AI or federated learning, which look at minimising memory consumption on individual devices, increasing privacy by keeping data on the local device, as well as reducing communication needs between distributed entities. Those features serve as the foundation for the use cases shown in Figure~\ref{fig:usecases}.

Based on the collective input of Dagstuhl Seminar on Edge Intelligence (21342), this paper aims to share an envisioned roadmap that can bring together key actors and enablers to further advance the domain of Edge AI. Sections \ref{sec:5G-beyond}, \ref{sec:future-cloud}, and \ref{sec:ai} cover the perspectives of 5G beyond, future cloud, and AI/ML, respectively. Section \ref{sec:roadmap} presents the envisioned road map and outlook.

\section{5G Beyond Perspective}\label{sec:5G-beyond}

The evolution of the fifth-generation networks towards the 6G era shapes the perspective of future networking. 
The developments along this path not only encompass network communication (e.g. speed, coverage, and resilience) but also quality and delay in computations.
To unlock the true potential of future networks with such a complicated structure, various technologies, at both hardware and software levels, need to coexist and cooperate. These include, for example, the creation of edge computing and communication fabrics or using self-learning technologies for dynamic network orchestration. Similarly, there are critical perspectives in holistic system trustworthiness, including security assurance mechanisms or confidential communication, computing, and learning. In this section, we elaborate on the challenges and opportunities.

\subsection{Communication and Computation with Human-in-the-Loop}

With the wide dissemination of smartphones and other personal carry-on devices, their significant computation, communication, and sensing capabilities become valuable to solve challenges in networking. The examples include local data acquisition as in federated learning, reducing the communication overhead as in device-to-device caching, and cooperating in executing computationally-intensive tasks. In future networks, mobile devices can decide if, when, where, and which fraction of a specific task to offload to a server at cloud or edge. Besides, they can take the role of computational worker, form pools of resources, and divide the tasks based on their preferences to optimize their utility and performance. 

While humans or human-driven devices may significantly contribute to Edge AI, such involvement raises several challenges. To model such challenges, one can use multi-agent systems. Subsequently, to address them, one can use various mathematical tools such as control and game theory. Furthermore, ML and AI play significant roles if there is some uncertainty and lack of information. In all of the above-mentioned steps, the specific characteristics of humans should be taken into account \cite{Maghsudi21:CMH}. In particular, humans often act based on heuristics and irrational influences, taking into account the factors such as social norms and peer pressure. When dealing with self-interested entities, it is essential to consider mutual trust and to respect the welfare of each entity \cite{Ye16:PBS}. Finally, using humans as a data source, e.g., using body sensors or GPS, proliferates privacy concerns that strongly couple with legal and ethical challenges. 

\subsection{Critical but Conflicting Actors and Applications}

In the transition from the current systems and networks to 5G beyond, the future needs of the societies become the driving force that creates use cases. As such, building innovative technology to address the society-driven use-cases becomes imperative \cite{matinmikko2020white}. To some extent, it stands opposite to the current use-cases such as low-latency and reliability that are generated by technological advances rather than taming directly from the society. Examples include the current vertical trends, including resource-efficient manufacturing, green energy generation and distribution, organic agriculture, and optimization of retail logistics.

Heterogeneous actors are expected to build and consequently share the massive edge computing infrastructure to serve their wide range of demands. Despite having some common goals to achieve, such actors often exhibit conflicting interests; i.e., more benefit for one attribute may reduce that of the other (here the utility can correspond to higher monetary return, sustainability, improved environmental factors, and the like). Finding a Pareto-optimal and stable solution to this problem is significantly challenging as different utility measures are often conflicting. The problem becomes aggravated in practice as it involves several decision-makers instead of a single central authority. That is not only because of self-interest but also due to information asymmetry and different types/preferences~\cite{Maghsudi19:DTM}. AI can be a solution to this problem, as it enables distributed systems to interact, learn, and make decisions -- rendering smart systems inseparable blocks of edge intelligence.  

\subsection{Edge Intelligence and the Emerging Technologies in Beyond 5G Networks}    

Next-generation networks beyond 5G encompass several technologies whose efficient deployment depends strongly on reliable cloud infrastructure as well as edge intelligence. These include, among others, joint communication and sensing, campus networks, Open Radio Access Network (RAN), intelligent reflecting surfaces (IRS), to name just a few. For example, the technology of joint communication and radar sensing is implementable in two networking architectures, namely small cell networks, and cloud RAN \cite{Rahman20:EJC}. While the latter is amenable to the cloud infrastructure, both implementations greatly benefit from edge intelligence. That is because joint communication and sensing necessitates swift signal processing and precise pattern recognition, both of which are computationally complex. 

Another example is campus networks, which covers a geographically limited region to cover the communication requirements specific to that area. For example, a manufacturing company can integrate a campus network in response to the need for secure, reliable, and persistent industrial communications with ultra-low latency. Other applications of campus networks include agriculture fields, construction sites, hospitals, and the like. The 5G technology, together with the edge computing capacity and AI, are the drivers of campus networks. They enable secure and stable communication, fast and no-failure computation, also reliable and efficient performance, even in the absence of precise information. The reliable performance of several other technologies such as IRS depends on edge intelligence as well. IRS technology relies on the optimal beam configuration, which might happen repeatedly. As a result, the required low-delay computation can be handled by the edge.  

\subsection{Technology Meets Business}

Recently, the discussion around 5G and beyond has been largely driven by the potential use cases and the over-arching goal to build real-time integrated edge computing, AI, and communication services that respond to the dynamic needs of the applications. These anticipated solutions, from everyday life to smart traffic and medical advantages, are significant but  need evaluations within the context of technology and novel business models. 

We identified three examples where AI technologies have a role in entirely new functions, however, hardly any business cases and models are yet defined for them: 
\begin{enumerate}
    \item \textbf{Interpreting the results of joint sensing and communication capabilities in future networks.} Future higher frequency communications allow some level of “radar-like” recognition of the environment. This represents a significant change to what current networks do or how they operate, and the security issues involved are substantial. 
    \item \textbf{Optimal link-level communication details discovery through ML.} This, in theory, is possible. However, the usefulness of these technologies is still questionable along with the necessary learning costs offset by the optimisation benefits accrued.
    \item \textbf{Interoperability and collaborative use of data and AI technologies.} Current systems are largely run within single organisations, but we raise the question: “what can we do to enable more sharing technologies and inter-operable interfaces”. There is a need for various stakeholders to identify and discuss relevant security, privacy, and ethical issues and tools to respond to them in trustworthy manner. 
 \end{enumerate}

\section{Future Cloud Computing Perspective}\label{sec:future-cloud}

%
The perspective of future cloud computing is shaped by the collaboration between future clouds and future edge nodes.
The main reasons for shifting tasks from the cloud to the edge are latency, bandwidth, locality of data, scalability, accessibility, security, and fault tolerance.
This collaboration is not only a technical aspect, but also involves business and stakeholder challenges.
An important non-technical challenge is the potential competition between cloud and edge, which need to work together closely to provide the best-possible service to customers.
In the following, we will go into detail about the technological challenges involved from the cloud perspective. In our opinion, these challenges are: i) resource management, ii) energy constraints and efficiency, iii) security, trust and privacy, and iv) intermittent connectivity.

\subsection{Resource Management}
Although the total volume of edge servers may provide a large amount of resources, the necessary locality of edge servers limits the amount of available resources compared to a cloud environment quite drastically.
That limits the number of different ML tasks that may be executed simultaneously due to hardware limitations and the execution latency of the tasks.
While the number of ML tasks supported by the edge can be increased through on-demand loading and execution of the corresponding models, this may increase the latency of tasks that need to be loaded.
Thus, for latency-critical tasks, pro-actively reserving resources might be a necessity.
As the reservation of resources could drastically limit the number of supported ML tasks, the necessity of reserving resources should be determined based on the risk of failure together with the consequences of that failure.
Especially when resources are sparse, the management of these resources in critical situations might be challenging.

Importantly, AI tasks at the edge are often embedded in larger settings, e.g., as part of a control workflow.
Accordingly, different services hosted on edge resources have to interact with each other.
Not taking into account that data items may have to be forwarded to services may lead to the fact that these services are not ready when receiving data and/or may not possess the computational resources for handling data items.
Accordingly, a backpressure of data items might occur.
This is especially the case in stream processing scenarios, e.g., when new data items need to be classified and---based on the classification---forwarded to different destination services.
Thus, it might be interesting to observe and analyze the execution of tasks depending on their priority level given different loads on the edge.

\subsection{Energy and Operational Constraints}
In general, edge nodes are expected to be less efficient (in terms of energy and cost) than cloud data centers. That is, as the Economies of Scale might work against edge data centers, e.g., by allowing for better cooling. Both, edge nodes and cloud data centers, have the possibility to deploy renewable energy, but it is unclear where these units are producing energy more efficiently. On the one hand, large data centers provide more opportunities to deploy renewable energy, increasing their sustainability and environment compatibility, but also consume a significantly higher amount of energy compared to edge nodes. On the other hand, edge nodes might harvest their own energy easier than cloud data centers due to their small scale and geographical dispersion, even though their capacity for renewable energy units is quite limited.

In addition, the data need to be transferred via the Internet which also consumes energy, increasing the possibility of edge nodes to surpass energy-efficient cloud data centers: Edge nodes can pre-process data, e.g., can detect anomaly cases and transmit only relevant data or adapt sensing rates dynamically. However, energy saving largely depends on the amount of processing that can be done at the edge. Another possibility is to move the training process partially to the edge, utilizing methods such as federated learning (see Sec.~\ref{sec:VerticalFederatedLearning}). In that case, only learned parameters (e.g., gradients) need to be transmitted, rather than raw data, which is multiple orders of magnitude larger. 

Another potential advantage of the edge nodes is the more specially designed hardware compared to generic cloud server racks, as this specially designed hardware is becoming more popular in the market (e.g., hardware accelerators) and utilizes limited processing and graphical capacities more efficiently.

\subsection{Security, Trust, and Privacy}
\begin{figure}
    \centering
    \includegraphics[width=\linewidth,trim= 4cm 3cm 4cm 3cm, clip]{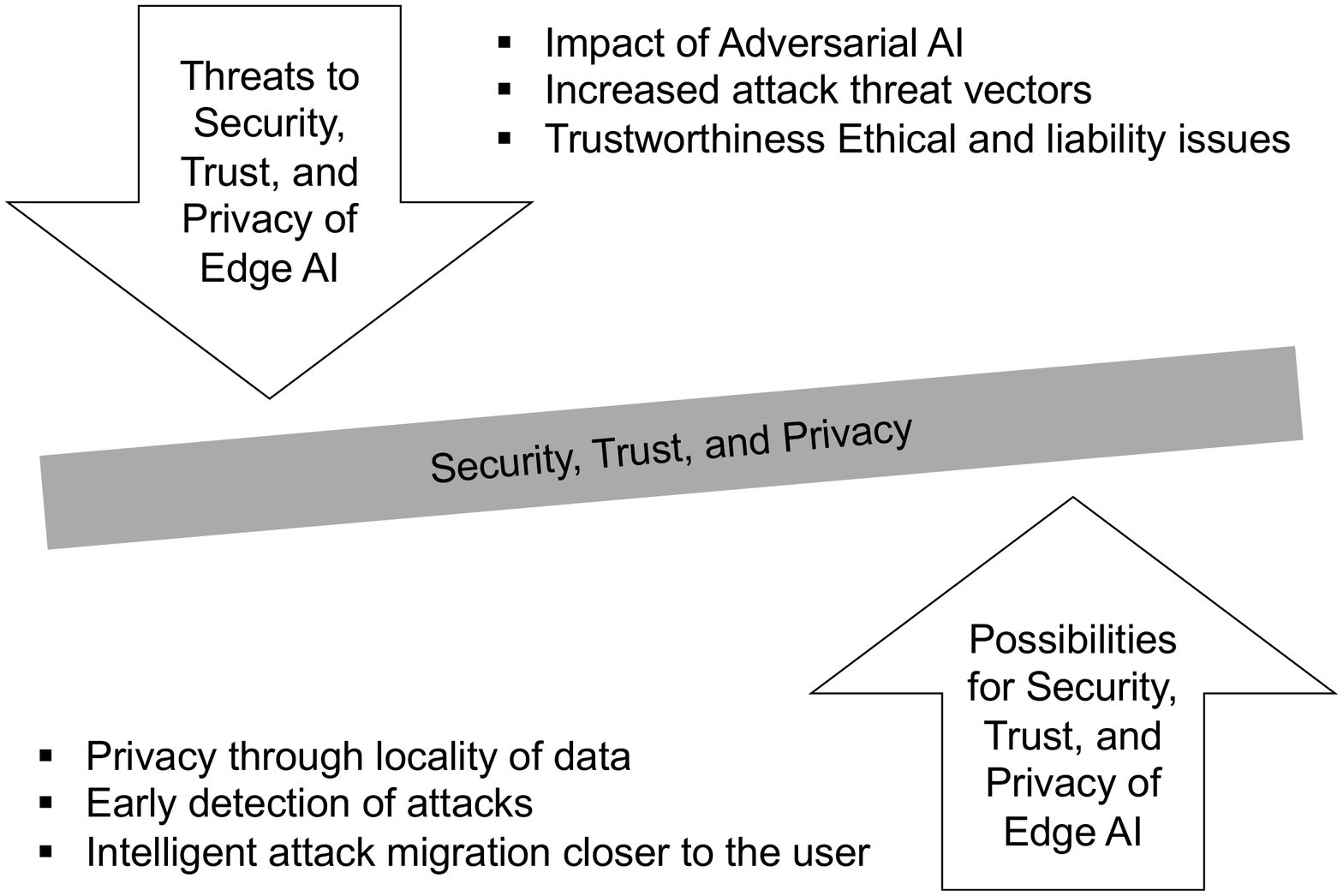}
    \caption{Trust, Security and Privacy of Edge AI}
    \label{fig:EdgeAI-SP}
\end{figure}

As shown in Figure~\ref{fig:EdgeAI-SP}, Edge AI offers both opportunities for higher trust, security, and privacy, but also adds additional challenges.
The security might be increased as some attacks might be detected early and countermeasures can be taken.
However, the number of possible targets and attack vectors are much higher, which lead to a higher potential of attacks.

Edge AI might have an issue with trust in the system, which is caused by its distributed nature and the potential liability issues.
However, the trust in edge servers might be increased through open implementation and specifications.
Additionally, it might be necessary to differentiate between devices under the control of the user and resources provided by providers (e.g., AWS, GCP, Azure).
One possible solution for increasing trust is the introduction of a reputation system known from other areas, that assesses and manages trust.
However, it will take new players a while to build that reputation.

Another property of Edge AI is that the data required to perform the ML tasks is kept local at the edge servers.
This led to the common assumption that the privacy and trust in edge servers is higher compared to the cloud.
In contrast to the common assumption, edge does not guarantee privacy, but raises new challenges to ensure privacy.
That is, as the trust of edge nodes is harder to keep and manage compared to a cloud provider, thus also aggravating data security and privacy, even though the locality of the data will prevent some attacks on the data.

\subsection{Intermittent Connectivity}
In general, we also consider the possibility of poorly connected edge servers, i.e., edge servers in regions where the connection to the internet is poor.
Also in those areas, edge intelligence has a huge potential and can be a driver towards digitalized non-urban areas.
Some examples for such systems include water and air pollution monitoring and natural disaster (wildfire, flood, volcanic eruption, etc.) prediction.
For instance, in water pollution monitoring, the identification of substances in the water could be done using edge resources.

The lack of reliable network connectivity in those areas could be the main driving factor in environmental monitoring scenarios in which streaming data has to be processed in near real-time.
Intermittent connectivity and energy constraints prevent monitoring systems from continuously transmitting raw data to the cloud for processing, whereas less data-intensive control signals like the output of ML algorithms can still be communicated under interruptions and low bandwidth availability.
In addition, the actors which rely on the outcome of ML algorithms are often close to the data sources.
By not sending the data to the cloud, a complete processing step can be saved, e.g., starting countermeasures automatically.
In the previous example of water pollution monitoring, the services running at the edge can then inform pumps or valves to open or close, depending on the scenario.

\section{Evolving AI/ML Perspective}\label{sec:ai}

\begin{figure}
    \centering
    \includegraphics[width=\linewidth,trim=0 3cm 0 4.8cm, clip]{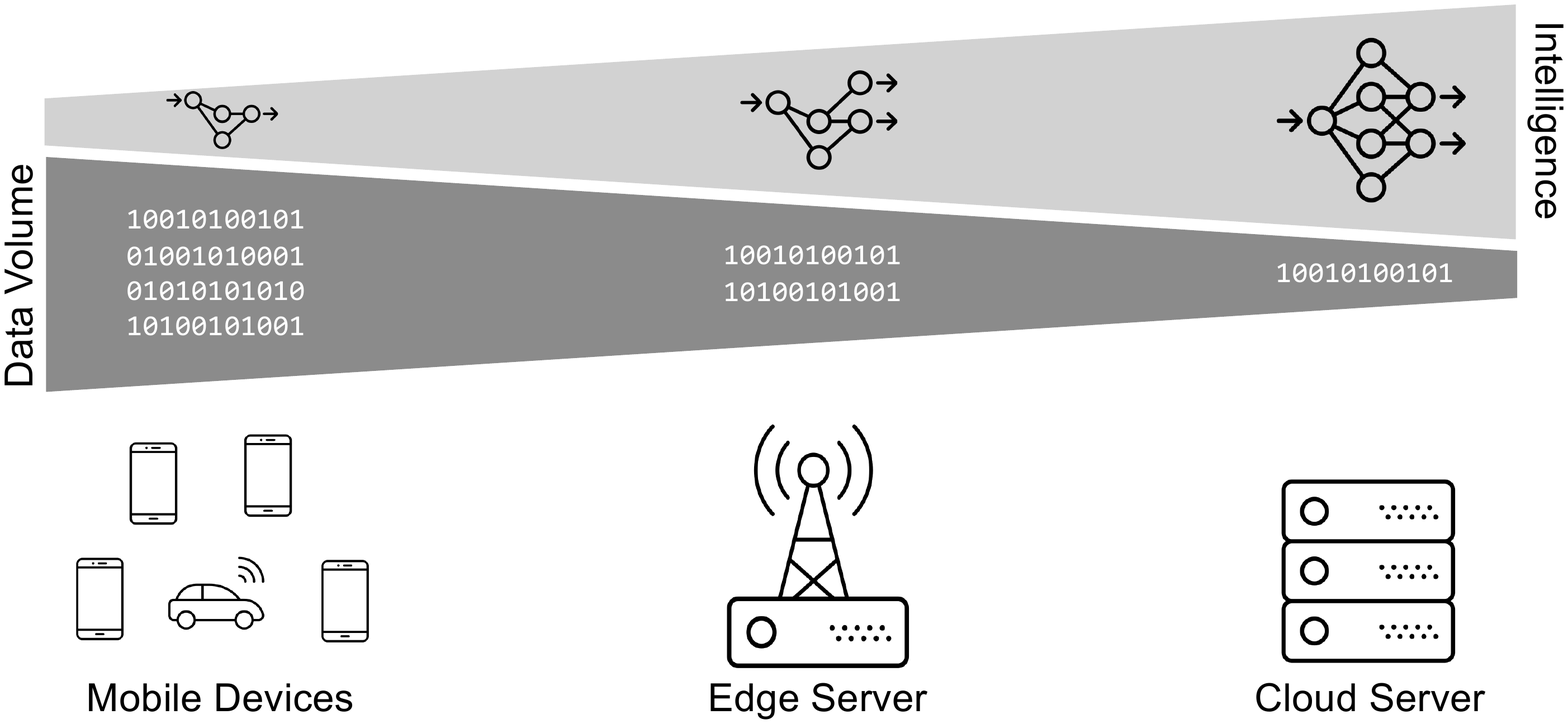}
    \caption{Scope of Edge AI}
    \label{fig:EdgeAI-scope}
\end{figure}

In this section, we describe the AI perspective on edge nodes and the challenges associated with it.
Figure \ref{fig:EdgeAI-scope} depicts the different levels of complexity of AI and the availability of data on mobile devices, edge servers, and cloud servers: While the mobile devices have the highest volume of data available to them, their capability of processing this data is generally limited.
Thus, preprocessed data can be offloaded to local edge servers to perform tasks which cannot be executed by the mobile devices.
The edge server has more resources than the mobile devices and can run more complex tasks, but due to the preprocessing its access to data is more  limited.
Tasks that cannot be executed will then again be offloaded to the cloud server, which can run the most complex models, but also receives only a part of the data available at the edge server.

In the following, we describe three challenges that have been identified and deserve community attention: 
These challenges are 1) availability of accelerators for AI applications; 2) defining a trade-off between accuracy and resource demand; and 3) utilising federated learning of edge servers.


\subsection{Accelerators for AI applications}
While we expect that hardware accelerators for AI applications will also be available at the edge, it is expected that only small hardware accelerators will be available there.
That is, as resources at the edge can be used less elastically compared to the cloud, which might lead to idle hardware.
Due to business aspect, it is expected that the large hardware accelerators will be deployed in the cloud.

One current trend to work with these limited resources is to split the model and AI algorithm into multiple smaller chunks, which can be executed by these small hardware accelerators.
Through joint interference among these small hardware accelerators, the user experience and quality of service of applications running at the edge can be enhanced.
In general, it will be necessary to transfer the large models with high resource demands to the cloud and vice-versa.

ML also provides unique opportunities for offloading and multi-tier architectures.
Unlike traditional workloads, which inherently offer conflicting requirements in terms of scaling to more nodes and tiers and keeping the state of the workload sufficiently consistent between the different nodes, ML, especially training of neural networks, has clearly defined state distribution and synchronisation models \cite{10.1145/3377454}.
Furthermore, unlike traditional distributed systems the consistency requirements can be comparably easily relaxed, thereby trading network bandwidth and number of replicas for typically moderate losses in accuracy. An important question is therefore which accuracy and latency requirements the application has and how to monitor the quality of the model during the entire life-cycle on the edge.

\subsection{Tradeoff between accuracy and resource demand}
As mentioned previously, edge servers have higher resource constraints compared to cloud servers. Some models may only run in the cloud due their complexity.

In today's AI research, the common goal is often to achieve results with the highest possible accuracy or maximizing the reward function.
However, a pivotal difference for Edge AI is that the quality of results is not the only measure for performance, but also other metrics like energy consumption and memory need to be considered.
Thus, models need to be elastic to adapt to the currently available resources and memory.
For that purpose, it must be possible to adapt the size of the respective ML models.
This can be done by miniaturizing them using techniques like transprecision, quantization or approximation.
These techniques generally provide a tradeoff between performance metrics and the accuracy of the model.
Examples for Edge AI minimization are TinyML and TensorLite.
An interesting research challenge is the analysis of adapting the ML model given the current (and maybe future) available resources.
In addition to that, we consider the possibility to offload certain tasks to the cloud: If the accuracy of the miniaturized model executed at the edge is insufficient, it might be necessary to offload the task to the cloud.
The cloud is then able to provide the results with high accuracy and send it back to the edge.

\subsection{Federated Learning}
\label{sec:VerticalFederatedLearning}
In addition to the miniaturization of models, models can be trained on multiple devices simultaneously using federated learning.
This is especially important for Edge AI, as the distributed nature of edge-servers increases the need of federated learning.

For federated learning, the data is kept local and thus not shared with the cloud.
However, when a complex model is run at an edge-server, the available resources might be insufficient to run or train the model.
One research challenge is the deployment of a complex model only in the cloud, which is trained in a federated manner based on less complex models deployed at the edge.

Federated learning also allows for the classification of time-critical labels at the edge directly, while other labels need to be classified in the cloud.
This adds another layer of elasticity to the model.
We expect that the accuracy of these time-critical labels remains high even though the model might be simplified through miniaturization by summarising all non-time-critical labels to be classified by the cloud model.
Thus, an open research question is the miniaturization a model such that the accuracy for a subset of labels remains high and the decision on when a classification needs to be executed by the cloud.

In addition, the data received by the edge servers and generated by the mobile devices might be quite diverse.
Federated learning can help to train a single model using these diverse sensor inputs and increase the accuracy of all models independent of their sensor setup.
The detection of malicious data, that could harm the training process, is also considered as an open challenge \cite{tolpegin2020data}.

Another challenge is the elastic adaptation of federated learning systems to dynamically updated client partnerships considering the satisfaction of collaboration criteria (e.g., minimum number of training partners) and still optimizing model accuracy.


\section{Roadmap and Outlook}\label{sec:roadmap}


\begin{figure*}
    \centering
    \includegraphics[width=\linewidth,trim=0.75cm 12cm 0.5cm 3.75cm, clip]{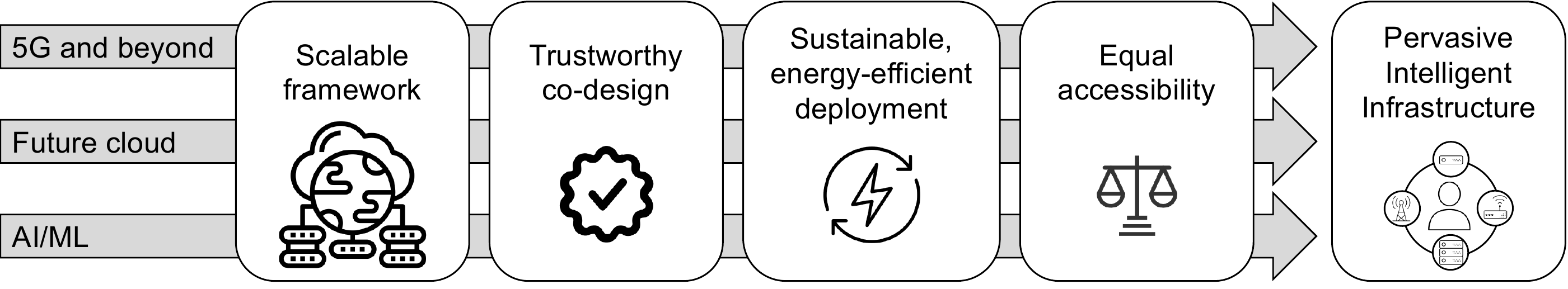}
    \caption{Roadmap of Edge AI}
    \label{fig:roadmap}
\end{figure*}

\textbf{Roadmap overview}: The envisioned roadmap of Edge AI is highlighted in Figure~\ref{fig:roadmap}. Along with three identified driving areas of 5G beyond, future cloud, and evolved AI/ML, the advancement of different technologies and the growing business interests will take Edge AI forward in terms of hardware, software, service models, and data governance.

Starting from the current state of play driven by cellular, cloud, and AI/ML service providers, the roadmap reflects five general phases: scalable framework, trustworthy co-design, sustainable and energy-efficient deployment, equal accessibility, and pervasive intelligent infrastructure. As changes can always occur, the sequence depicted in the roadmap could be switched or combined. Nonetheless, this Edge AI roadmap reflects the combined effects of technology enablers and non-tech demands such as socioeconomic transformation of user behaviours, purchasing power and business interests. 

\textbf{Open research challenges}: Despite of its promise and potential, Edge AI can face major challenges in large scale deployment, including energy optimization, trustworthiness, security, privacy and ethical issues.

As an important goal of sustainability, the energy consumption of Edge AI needs to be optimized. The energy efficiency is crucial for Edge AI embedded infrastructures (e.g., road side units, micro base stations) to sustainably support advanced autonomous driving and Extended Reality (XR) services in the years to come. Through the pipeline of data acquisition, transfer, computation, and storage, there exists the possibility for Edge AI to trade accuracy with reduced power and less time consumed. For instance, noisy inputs from numerous sensors can be selectively processed and transferred in order to save energy. A set of applications would be satisfied with an ‘acceptable’ accuracy instead of exact and absolutely correct results. By introducing this new dimension of accuracy to the optimization design, the energy efficiency can be further improved.

Concerning trustworthiness, Edge AI benefits from its close proximity to the end-devices. However, due to the distributed deployment with deep insights into personal context, the safety and perceived trustworthiness for Edge AI services are raising concerns among the stakeholders (e.g., end users, public sectors, ISP)~\cite{Toussaint-CogMI2020, Toussaint:SenSys2021}. To achieve trustworthy Edge AI, critical building blocks are needed, including verification and validation mechanisms that ensure transparency and explainability, especially in the training and deployment of Edge AI in decentralized, uncontrolled environments. The trustworthiness of Edge AI is a stepping stone to establish an appropriate governance and regulatory framework, on which the promise of Edge AI can be built.

\textbf{Safety and Privacy/Ethical Issues}: When we discuss the security and privacy of Edge AI, there are two aspects to consider, i.e., 1) usage of Edge AI to provide security and privacy; 2) new security and privacy issues due to the use of Edge AI, as in Figure~\ref{fig:EdgeAI-SP}. Edge AI can be a vital tool to ensure the security and privacy network. Instead of performing AI processing in the cloud, Edge AI processes the data locally closer to the user. That prevents the necessity of transferring data between the user and the cloud and eliminates the possibility of attack during the backhaul data transmission phase. Moreover, privacy can be improved by keeping and processing data locally. Processing is focused on moving the interface of the AI workflow to the device while maintaining data  restricted  to  the  device. Another security benefit of Edge AI is the possibility of an AI algorithm being decentralized and eliminating a single point of failure in the cloud AI system. In a way, Edge AI could be a reason to limit the impact of an attack on the  local environment and mitigate it already at the edge level due to added intelligence. There are new safety and privacy issues arising with the use of Edge AI. Decentralisation of Edge AI opens up unknown attack vectors and increases the number of entry points for AI system attackers. Edge AI devices might not have the same level of security as cloud, which can be used as easy entry points to get access to attack the AI system. Moreover, edge devices are physically accessible, and added AI to edge devices makes the impact of captures edge devices severer than an edge device without AI. For instance, taking control of Edge AI devices might jeopardise or take control of almost all the localised network services of a particular location.

\textbf{Concluding remarks}: The promises of Edge AI come hand in hand with new challenges and uncertainties. This paper is our endeavour to capture the latest technological development, crucial actors across three major dimensions of 5G-beyond, future cloud and AI/ML for the envisioned roadmap. We hope that the perspectives conveyed in this paper can provide a different view to the community and further pave the way to promote the global rollout of Edge AI in the long run.

\begin{acks}
	The discussions leading to this editorial were initiated in
	Dagstuhl Seminar 21342 on
	\emph{Identifying Key Enablers in Edge Intelligence},
	and we thank all participants for their contributions. 
	The work is partially supported by the European Union's Horizon 2020 research and innovation programme under grant agreement No. 101021808, by CHIST-ERA grant CHIST-ERA-19-CES-005, and by the Austrian Science Fund (FWF): I 5201-N.
\end{acks}

{ 
{
    \bibliographystyle{ACM-Reference-Format}
    \bibliography{bibliography.bib}
}
}

\end{document}